# Vortices and antivortices in antiferroelectric $PbZrO_3$

Ying Liu[1,2,†], Huazhang Zhang[3,†], Konstantin Shapovalov[3†], Ranming Niu[2], Julie M. Cairney[2], Xiaozhou Liao[2], Krystian Roleder[4], Andrzej Majchrowski[5], Jordi Arbiol[1,6], Philippe Ghosez[3], Gustau Catalan[1,6]

[1]Catalan Institute of Nanoscience and Nanotechnology (ICN2), Campus Universitat Autonoma de Barcelona, Bellaterra 08193, Catalonia.

[2]School of Aerospace, Mechanical and Mechatronic Engineering, and Australian Centre for Microscopy and Microanalysis, The University of Sydney, Sydney, NSW 2006, Australia

[3]Theoretical Materials Physics, Q-MAT, Université de Liège, B-4000 Sart-Tilman, Belgium

[4] Institute of Physics, University of Silesia, Ulica 75 Pułku Piechoty 1, 41-500 Chorzów, Poland

[5] Institute of Applied Physics, Military University of Technology, ul. Kaliskiego 2, 00-908 Warsaw, Poland

[6]Institut Català de Recerca i Estudis Avançats (ICREA), Barcelona 08010, Catalonia.

[†]These authors contribute equally to this work.

**Abstract**

Although ferroelectric materials are characterised by their parallel arrangement of electric dipoles, in the right boundary conditions these dipoles can reorganize themselves into vortices, antivortices and other non-trivial topological structures. By contrast, little is known about how (or whether) antiferroelectrics, which are materials showing an antiparallel arrangement of electric dipoles, can exhibit vortices or antivortices. In this study, using advanced aberration-corrected scanning transmission electron microscopy, we uncover the existence of atomic-scale (anti)vorticity in ferroelastic domain walls of the archetypal antiferroelectric phase of $PbZrO_3$. The finding is supported, and its underlying physics is explained, using both second-principles simulations based on a deep-learning interatomic potential, and continuum field modelling. This discovery expands the field of chiral topologies into antiferroelectrics.

Vortices, antivortices and related topological structures such as skyrmions and hopfions have been of interest in both ferromagnetic and ferroelectric materials for the past few decades[1, 2, 3, 4, 5, 6, 7, 8, 9, 10, 11, 12]. Their singular structure combines small size and topological protection, rendering them potentially useful for applications in high-density data storage[13] and logic devices[14]. In addition, ferroelectric thin films with vortex domains have been reported to display negative capacitance, with the potential to reduce energy consumption in transistors[15], while enhanced electric conductivity was observed at ferroelectric vortex cores in $BiFeO_3$[16].

The driving force for the appearance of vortices in ferroelectric materials is the need to minimize the depolarizing fields that appear at discontinuities perpendicular to the spontaneous polarization, such as the interfaces between polar and non-polar materials in ultra-thin films and superlattices[5, 6, 7, 17, 18]. When the polarization curls into itself, polar continuity is preserved (div **P**=0) and depolarization fields vanish[6, 7, 11, 12, 18]. By contrast, in antiferroelectric materials, adjacent dipoles are antiparallel, leading to an overall polarization of zero[19]. Antiferroelectrics are thus expected to be free from depolarizing fields and lack the drive to form vortices or antivortices. It is therefore natural to wonder whether antiferroelectrics can produce such topological structures. Here we report that they can.

The material chosen for this study is $PbZrO_3$, which was the first discovered antiferroelectric material, and is considered an archetype of antiferroelectricity[20]. It displays at room temperature an ↑↑↓↓ electric dipole arrangement within its unit cell[21]. The room temperature crystal structure is orthorhombic (space group *Pbam*[22] and can display up to 6 ferroelastic domain orientations[23]. Ferroelastic domains appear in crystals whose spontaneous strain tensor (with respect to a higher-symmetry paraelectric phase) can have different energy-equivalent orientations. Many ferroelectrics are also ferroelastic, and in those the polarization rotates across the wall keeping the polar component perpendicular to the wall constant to minimize electrostatic energy cost (again, div **P**=0), although charged configurations with head-to-head or tail-to-tail dipoles sometimes appear[24, 25, 26, 27]. In antiferroelectric $PbZrO_3$, electron microscopy studies[28] suggest that the antiparallel dipoles rotate across ferroelastic walls in a similar way as in a ferroelectric. As we shall see, however, these polar rotations are not homogeneous, and at the locum of the domain wall they develop into arrays of vortices or antivortices. We show two different examples in Figure 1, which displays two domain walls where the antipolar axis rotates by 90 degrees across the wall.

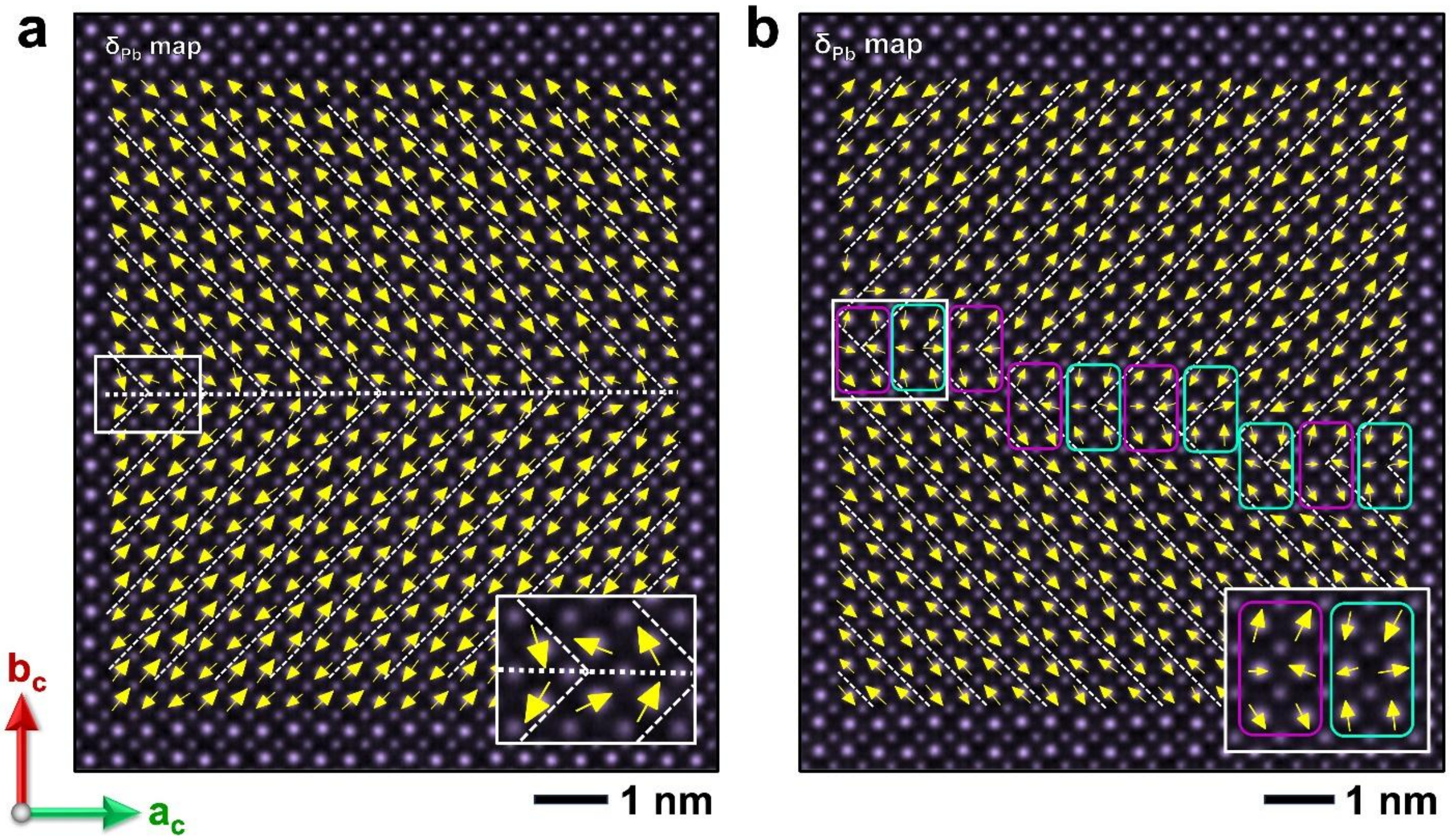

**Fig. 1 |** STEM-HAADF image acquired from straight twinning boundaries and overlaid by corresponding Pb displacement maps. (a) Head-to-tail Pb displacement configurations at 90° twining boundaries in $PbZrO_3$ single crystal. The inset in (a) shows a magnified view of the rectangular region outlined at the bottom of the image. (b) Head-to-head and tail-to-tail Pb displacements and concomitant appearance of antivortices were also observed, highlighted by turquoise and purple rectangles, respectively. The inset in (b) shows a zoomed-in view of the section marked by a white rectangle at the image's lower part.

Figure 1 presents atomic resolution scanning transmission electron microscopy (STEM) high angle annular dark field (HAADF) images of 90° twining boundaries in a $PbZrO_3$ single crystal. In these images, the Pb columns appear as bright spots, while the Zr columns are dimmer. To reveal the dipole configuration, we extracted the Pb displacement map ($\delta_{Pb}$, yellow arrows, indicating lead displacements relative to the centre of their four nearest Zr atoms) using a Python-based peak finding method (see more details in Refs[29, 30, 31, 32]). This is represented by the yellow arrows. Ferroelastic 90° domain walls are revealed and delineated by a white dotted line in Figure 1a. In the first wall (Figure 1a), the dipoles have a head-to-tail configuration across the wall, so the perpendicular component of the polarization is continuous, and depolarization effects are not expected. Nevertheless, we observe that the dipoles directly adjacent to the wall are somewhat off their ideal structural positions, already showing a tendency towards vorticity (see inset with a close-up). As we shall see, this faint vorticity is indeed expected at the twin boundaries. Meanwhile, when the local dipole configuration at the ferroelastic wall is head-to-head or tail-to-tail (Figure 1b), there is a polar discontinuity and depolarization effects may be expected. Correlated with this polar discontinuity, well-defined topological antivortices (winding number = -1 as per Mermin's definition[33]) are observed.

These observations prompt several questions: What is the origin of these vortices and antivortices? Does depolarization play a role in their emergence? Given that we have seen two different polar configurations at the walls, which one is energetically more favourable? can there be other configurations? To address these questions, we turn to theory, starting with second-principles atomistic simulations of $PbZrO_3$.

Based on the differences in the twinning planes (PbO plane and $ZrO_2$ plane) and the relative positions along the $a_c$ (with shifts of zero, one, two, and three cubic unit cells) and $c_c$ directions (with no shift and a shift of one cubic unit cell) between the two sides of the twinning boundaries, we constructed 16 initial model configurations (Supplemental Materials Figure S1). Their structures were then relaxed using second-principles simulations and converged to 8 different configurations, as detailed in Supplementary Materials Figures S2 to S5. Figure 2 presents four typical relaxed twinning boundary structures. Figures 2a and 2b illustrate the head-to-tail domain configurations for the PbO and $ZrO_2$ twinning planes, respectively, aligning with the experimental results previously shown in Figure 1 and Supplementary Materials Figure S6. Figures 2c and 2d depict the head-to-head and tail-to-tail Pb displacement configurations, in which both vortex-like and antivortex patterns are evident, as observed experimentally (see also Figure 4-a).

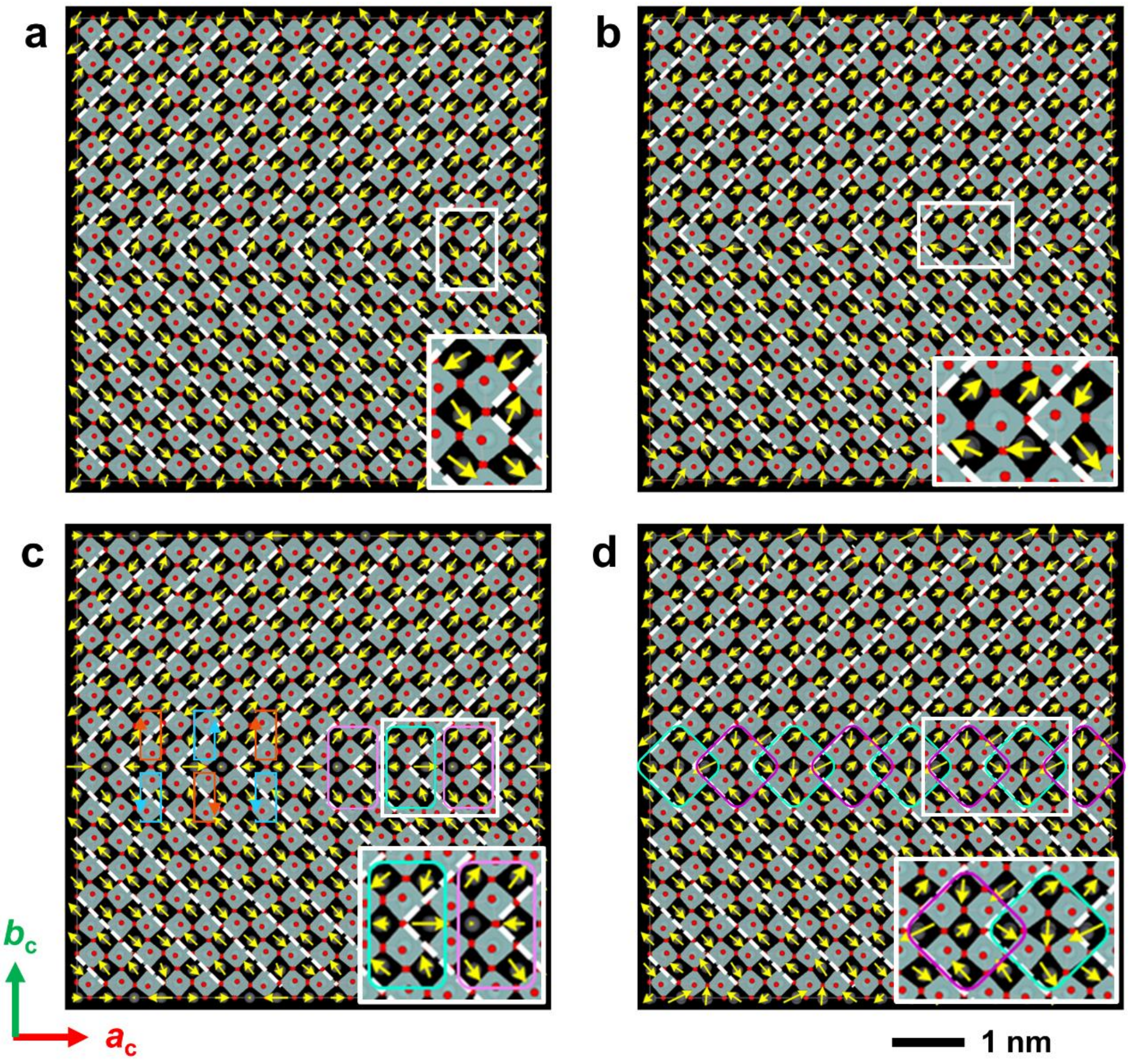

**Fig. 2 |** Four typical twinning boundary configurations obtained from second-principles relaxation. (a) A head-to-tail twinning boundary that has its twinning plane located at the PbO plane, and accompanied by a smooth rotation of Pb displacement. (b) A head-to-tail twinning boundary exhibits a smooth rotation of the Pb displacement vector and has its twinning plane situated on the $ZrO_2$ plane. (c, d) Head-to-head and tail-to-tail twining boundaries characterized by vortices-like (rectangular arrows in blue and orange) and antivortices domain configurations (rectangles in purple and turquoise).

The relative stability of different types of twinning boundaries was evaluated by comparing the domain wall energies. The data for all eight types of twinning boundaries can be found in Supplementary Materials, Table S2. From these, we have selected three representative cases that are shown in figure 2: two neutral walls centered on the PbO plane and the ZrO2 plane (figures 2a and 2b, respectively) and a charged (head-to-head) wall centered on the PbO plane, with coherent octahedral tilts across the wall. All other walls had higher energies and are less likely to be observed. The outcomes for the three types of twinning boundaries shown in Figure 2 are detailed in Table 1. The main conclusions that emerge from these calculations are that (i) head-to-tail configurations, coherent octahedral rotations across the plane, and being centered at the PbO plane are all factors that reduce the domain wall energy, with the lowest-energy wall being the one that meets all these criteria, and (ii) whenever there are head-to-head/tail-to-tail configurations, antivortex-like structures appear, consistent with the need to reduce the polar discontinuity.

**Table 1.** Second-principles calculated total energies and domain wall energies for the four types of twinning boundaries depicted in Figure 2.

| | Type 1 **(Figure 2a)** | Type 2 **(Figure 2b)** | Type 4 **(Figure 2c)** | Type 8 **(Figure 2d)** |
|---|---|---|---|---|
| Domain wall energy ($mJ/m^2$) | 44.2 | 51.2 | 59.4 | 85.2 |

Up to this point, we have discussed possible vorticity, antivorticity and winding number at twin boundaries in the context of discrete systems. However, from a mathematical point of view, these concepts are only unambiguously defined for continuous fields rather than discrete (atomic scale) systems. In order to properly quantify the topological winding number, and to describe phenomenologically the origin of vorticity and antivorticity at the twin boundaries, we conduct a continuum-field study of $PbZrO_3$. For this, we use a 2D Landau energy model that has been used to successfully describe the stable modulated state of $PbZrO_3$ resulting in the ↑↑↓↓ antipolar displacement pattern [34]. As the *Pbam* phase of $PbZrO_3$ is also characterized by a tilting pattern of oxygen octahedra ($a^-a^-c^0$ in Glazer notation), the model takes into account its emergence and its interaction with polarization. In our study, polarization $\mathbf{P} = (P_x, P_y)$ and oxygen octahedral tilt $\boldsymbol{\phi} = (\phi_x, \phi_y)$ are described as continuum fields defined in $(x, y)$ coordinate space, with the Landau energy having the following form:

$$F = F_{\text{hom}}\ (\mathbf{P}, \boldsymbol{\phi}) + \frac{G}{2}\ (\nabla\mathbf{P})^2 + \frac{D}{2}\ (\nabla\boldsymbol{\phi})^2 - W\mathbf{P}(\nabla\boldsymbol{\phi})\boldsymbol{\phi}, \qquad (1)$$

where $F_{\text{hom}}\ (\mathbf{P}, \boldsymbol{\phi})$ is the homogeneous part of the energy described by an 8th order polynomial (see detail in Ref [34]), and G, D and W are the 4th rank tensors describing, respectively, polarisation correlation, tilt correlation and rotopolar gradient coupling – the full expression of the gradient terms and the numerical values of the corresponding tensor components are listed in Supplemental Information. Coordinate axes $x$ and $y$ are along the pseudocubic $\langle 100 \rangle$ directions, corresponding to $a_c$ and $b_c$ directions. Figure 3 shows the distribution of the polarization field $\mathbf{P}(x, y)$ obtained in head-to-tail (stable in our calculations) and head-to-head (metastable) twin boundaries, with the colour hue and the intensity corresponding, respectively, to the local direction and the magnitude of **P** (the corresponding distributions of the tilt $\boldsymbol{\phi}(x, y)$ are reported in the Supplemental Materials Figure S8). To allow an easier comparison with the experimental and the second-principles data, we overlay a vector map of the local polarization at a square grid corresponding to the Pb atomic sublattice.

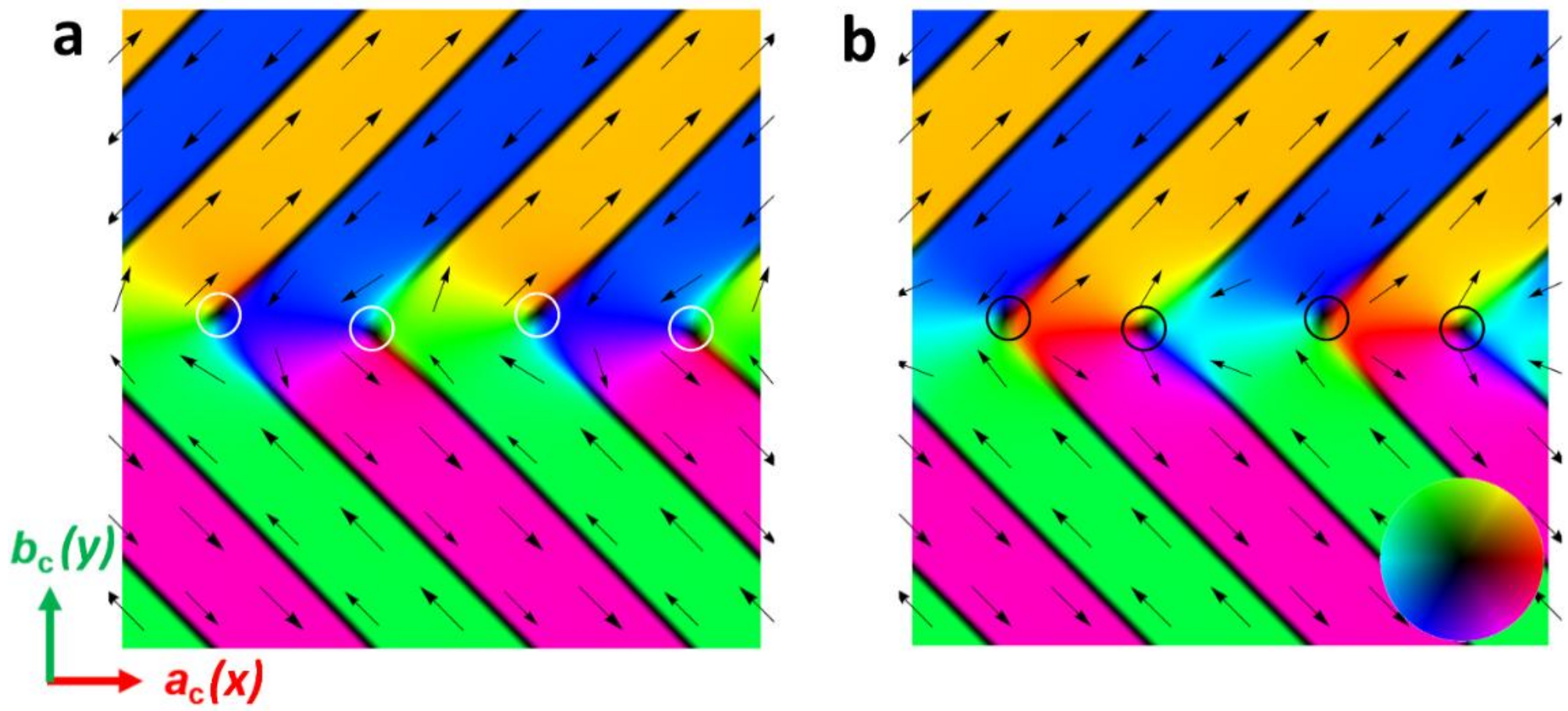

**Fig 3 |** Distribution of polarisation $\mathbf{P}(x, y)$ obtained in the continuum field study of twin boundaries in $PbZrO_3$: head-to-tail (a) and head-to-head (b). Colour hue and intensity correspond to, respectively, local orientation and magnitude of polarisation. White circles mark vortex cores with winding number $+1$, black circles mark the cores of antivortices with winding number $-1$. Inset: colour hue-intensity wheel.

According to Mermin's definition [33], the topological winding number is defined by the total change of the orientation angle of the vector field when going around a singularity (measured in increments of 2π). For a non-trivial topology of the polarization field, the two necessary conditions are (i) the existence of a **P**=0 singularity (the core of the vortex or antivortex), and (ii) that the winding number of the polarization around the singularity be different from 0 (+1 for vortices, -1 for antivortices). At the level of continuum fields, head-to-head walls obtained in our continuum-field study indeed display well-defined antivorticity, with antivortex cores ($\mathbf{P} = 0$) located at the twin boundary and winding number $-1$ around the cores. At the head-to-tail boundaries, vorticity is more questionable: vortex cores with $\mathbf{P} = 0$ and winding number $+1$ exist, but the "cores" are not fully detached from the antiferroelectric 180-degree wall steps. Since the integration path around the "vortex cores" contains a discontinuity, the change of angle is ill-defined. Our view from this analysis is therefore that the distorted polarization at neutral (head-to-tail) walls might not amount to true topological vortices, whereas the antivortices of charged (head-to-head) walls are topologically unambiguous.

In other ferroelectric systems, the formation of topological structures has been linked to depolarization fields [35] and/or to inhomogeneous strain [36]. Our results show clearer (anti)vorticity in the charged head-to-head/tail-to-tail ferroelastic walls compared to the neutral head-to-tail ones, suggesting indeed a role in minimizing depolarization. But the model also uncovers an additional role of the polarisation gradient term. The polarisation gradient term essentially imposes a smooth variation of polarisation across a boundary with a thickness given by the correlation length. The correlation lengths in our model of $PbZrO_3$ are: $\xi \sim 8$ Å at the 90-degree ferroelastic wall, and $\xi \sim 2$ Å at the antipolar 180-degree wall-like boundaries, and polar rotation emerges geometrically when these two intersect. We can show this by using a simplified continuum-field model taking into account only the polarisation-dependent parts of the Landau energy, i.e., only the first two terms of Eq. (1). Such simplified potential readily generates polar vortices and antivortices (see Supplemental Materials Figure S9).

Finally, we examine the dynamics. Here we highlight two important issues. First: if the head-to-head configuration costs more energy (as indicated by the second-principles calculations), ferroelastic walls should spontaneously evolve towards the low-energy head-to-tail configuration. Second: geometry alone dictates that the motion of a ferroelastic domain wall will switch between head-to-head and head-to-tail configurations, and the appearance and disappearance of antivortices is thus an inherent feature of ferroelastic domain wall motion in antiferroelectrics. Both predictions are experimentally confirmed (Figure 4).

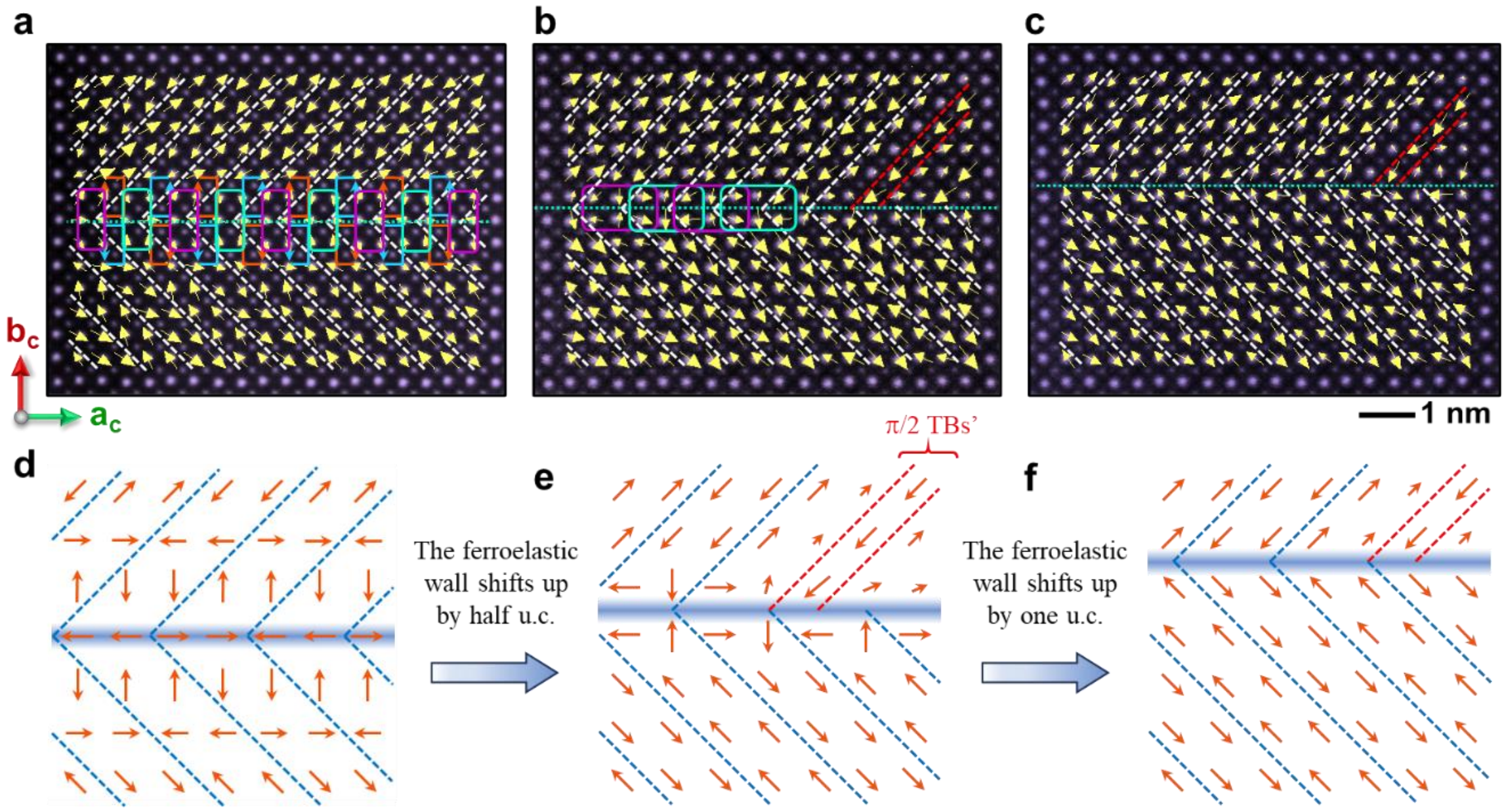

**Fig 4 |** Evolution of the domain structure under an electron beam. (a) A STEM-HAADF image overlaid with the corresponding $\delta_{Pb}$ map. The image was acquired in a $PbZrO_3$ thin film. The domain wall was created due to ferroelastic domain switching induced by electron beam irradiation[37]. Vortex-like and antivortex domains are indicated by rectangular arrows (in blue and orange) and rectangles (in purple and green), respectively. (b – c) Images of the same domain wall taken 8 days later. (d – f) Schematic diagrams of the domain evolution corresponding to (a – c). In (d – e), blue shaded areas indicate the ferroelastic domain wall; red arrows: Pb displacement; TB: twinning boundary; TBs': translational boundaries[29].

Figure 4 shows the evolution of a charged twin boundary under electron beam irradiation in TEM. Figure 4a displays a STEM-HAADF image overlaid with the $\delta_{Pb}$ map, highlighting the twin boundary configuration that emerged in the initial stage. Figure 4a shows an initially charged (head-to-head and tail-to-tail) configuration of the ferroelastic wall, with the coexistence of vortices (indicated by orange and blue rectangular arrows) and antivortices (green and purple rectangles). The PbO plane acts as the twinning boundary in this instance. Figure 4b displays the same twinning boundary, but captured eight days later. The domain wall has shifted by several tens of nanometers compared to Figure 4a. Figure 4c was captured just 1 minute after Figure 4b. The $\delta_{Pb}$ vector at the twinning boundary has altered, and the twinning boundary now exhibits a head-to-tail configuration with a smooth $\delta_{Pb}$ rotation. Outside the twinning boundary region, there is no significant change in the $\delta_{Pb}$. The $\delta_{Pb}$ configuration did not exhibit any further changes afterwards. As can be seen, the transition from head-to-head polarization in Figure 4b to head-to-tail polarization in Figure 4c has been achieved just by shifting up the position of the ferroelastic wall by one pseudocubic lattice unit, as expected from purely geometrical considerations (see Supplementary materials Figure S10). Further atomic resolution HAADF STEM images are provided in Supplementary Materials Figures S11 and S12.

The discovery of vortex and antivortex domains at ferroelastic twin boundaries provides new insights into the interplay between ferroelasticity and topology in antiferroelectrics. The combination of depolarization fields and polarization-gradient terms favours the formation of flux-closure polarization pathways, leading to atomic-scale antivorticity. The small (1 nm) scale of the observed structures sets a new benchmark for topological defects in crystals. Moreover, we see that as the ferroelastic wall moves through the lattice, it alternates between charged and neutral positions, raising the possibility of radiative effects from this alternation. More generally, the discovery of vortices and antivortices in PZO opens up antiferroelectrics to the wider field of topological materials research.

**Acknowledgements**

We thank J. Hlinka for insights on DW motion in antiferroelectrics. This work has been funded by the FET-Open programme of the EU, project TSAR (Topological Solitons in Antiferroics. Y. L. acknowledges the BIST Postdoctoral Fellowship Programme (PROBIST) funded by the European Union's Horizon 2020 research and innovation programme under the Marie Sklodowska-Curie Grant Agreement No. 754510. The ICN2 is supported by the Severo Ochoa Centres of Excellence programme, Grant CEX2021-001214-S. GC acknowledges the support of the Departament de Recerca i Universitats of Generalitat de Catalunya for the grants 2021 SGR 01297 and 2021 SGR 00457. KR's work was supported by the National Science Centre, Poland, Grant No. 2022/47/B/ST3/02778. The authors are grateful for the scientific and technical support from the Australian Centre for Microscopy and Microanalysis (ACMM) as well as the Microscopy Australia node at the University of Sydney.

**Data availability statement**

The second-principles atomistic simulations relies on the use of open-source packages such as ABINIT, DeepMD and LAMMPS (see SI).

The data files corresponding to the TEM images are available from the authors upon reasonable request.

Supplemental Materials for

# Vortices and Antivortices in Antiferroelectric $PbZrO_3$

Ying Liu[1,2,†], Huazhang Zhang[3,†], Konstantin Shapovalov[3†], Ranming Niu[2], Julie M. Cairney[2], Xiaozhou Liao[2], Krystian Roleder[4], Andrzej Majchrowski[5], Jordi Arbiol[1,6], Philippe Ghosez[3], Gustau Catalan[1,6]

[1]Catalan Institute of Nanoscience and Nanotechnology (ICN2), Campus Universitat Autonoma de Barcelona, Bellaterra 08193, Catalonia.

[2]School of Aerospace, Mechanical and Mechatronic Engineering, and Australian Centre for Microscopy and Microanalysis, The University of Sydney, Sydney, NSW 2006, Australia

[3]Theoretical Materials Physics, Q-MAT, Université de Liège, B-4000 Sart-Tilman, Belgium

[4] Institute of Physics, University of Silesia, Ulica 75 Pułku Piechoty 1, 41-500 Chorzów, Poland

[5] Institute of Applied Physics, Military University of Technology, ul. Kaliskiego 2, 00-908 Warsaw, Poland

[6]Institut Català de Recerca i Estudis Avançats (ICREA), Barcelona 08010, Catalonia.

[†]These authors contribute equally to this work.

**This PDF file includes:**

Second-principles computational details

Continuum-field calculation details

Figures S1 – S12

Tables S1 – S2

References S1 – S6

## Second-principles computational details

The second-principles calculations of the elastic domain walls in PZO were performed based on a previous developed deep learning interatomic effective potential [S1]. The model was trained using the DeePMD-kit [S2] and based on a first-principles dataset containing approximately $1.2\times10^4$ configurations calculated using ABINIT software [S3] and PBEsol functional [S4]. The model has been extensively validated, and found to be capable to accurately reproduce a series of (meta)stable phases in PZO. Here, we show in Table S1 the comparisons of the structural parameters and total energy for the *Pbam* phase between by the DFT calculation and the model-based calculation. For the detailed validations of the model we will report elsewhere.

Table S1. Structural parameters and total energy (with respect to the cubic reference structure) for the *Pbam* phase of PZO from the first-principles calculation using the PBEsol functional and the second-principles model-based calculation.

| | DFT (PBEsol) | | | Second-principles model | | |
|---|---|---|---|---|---|---|
| | *a* | *b* | *c* | *a* | *b* | *c* |
| Lattice parameter (Å) | 5.8742 | 11.7682 | 8.1723 | 5.8760 | 11.7663 | 8.1724 |
| | *x* | *y* | *z* | *x* | *y* | *z* |
| Pb1 (4g) | 0.19869 | 0.37642 | 0 | 0.19904 | 0.37575 | 0 |
| Pb2 (4h) | 0.20819 | 0.37146 | 0.5 | 0.20772 | 0.37108 | 0.5 |
| Zr1 (8i) | 0.74137 | 0.37572 | 0.75032 | 0.74158 | 0.37591 | 0.75051 |
| O1 (4g) | 0.77637 | 0.34256 | 0 | 0.77749 | 0.34211 | 0 |
| O2 (4h) | 0.80128 | 0.40573 | 0.5 | 0.80108 | 0.40509 | 0.5 |
| O3 (8i) | 0.53307 | 0.23786 | 0.71873 | 0.53362 | 0.23777 | 0.71925 |
| O4 (4f) | 0 | 0.5 | 0.20216 | 0 | 0.5 | 0.20259 |
| O5 (4e) | 0 | 0 | 0.22925 | 0 | 0 | 0.22851 |
| Total energy (meV/f.u.) | −285.9 meV/f.u. | | | −286.3 meV/f.u. | | |

The elastic domain walls of PZO in the *Pbam* phase parallel to the $(100)_{pc}$ plane was modeled in 16×16×2 supercells. Based on the different choices of twinning planes (PbO plane or $ZrO_2$ plane) and the relative positions along the *x* and *z* directions between the two domians, 16 initial structures were built and relaxed (Figure S3). The structural relaxations were performed based on the conjugate gradient algorithm as implemented in Lammps [S5]. The domain wall energy $E_{DW}$ was calculated according to

$$E_{DW} = \frac{E_{tot}-E_{pbam}}{2A_{DW}}, \quad (1)$$

where the $E_{tot}$ is the energy of the relaxed structure containing ferroelastic domain walls, $E_{Pbam}$ is the energy of the *Pbam* phase in the same size of supercell, $A_{DW}$ is the area of each domain wall in the relaxed structure, and a prefactor 2 is used because the simulation structures contains 2 domain walls so as to fulfill the periodic boundary conditions.

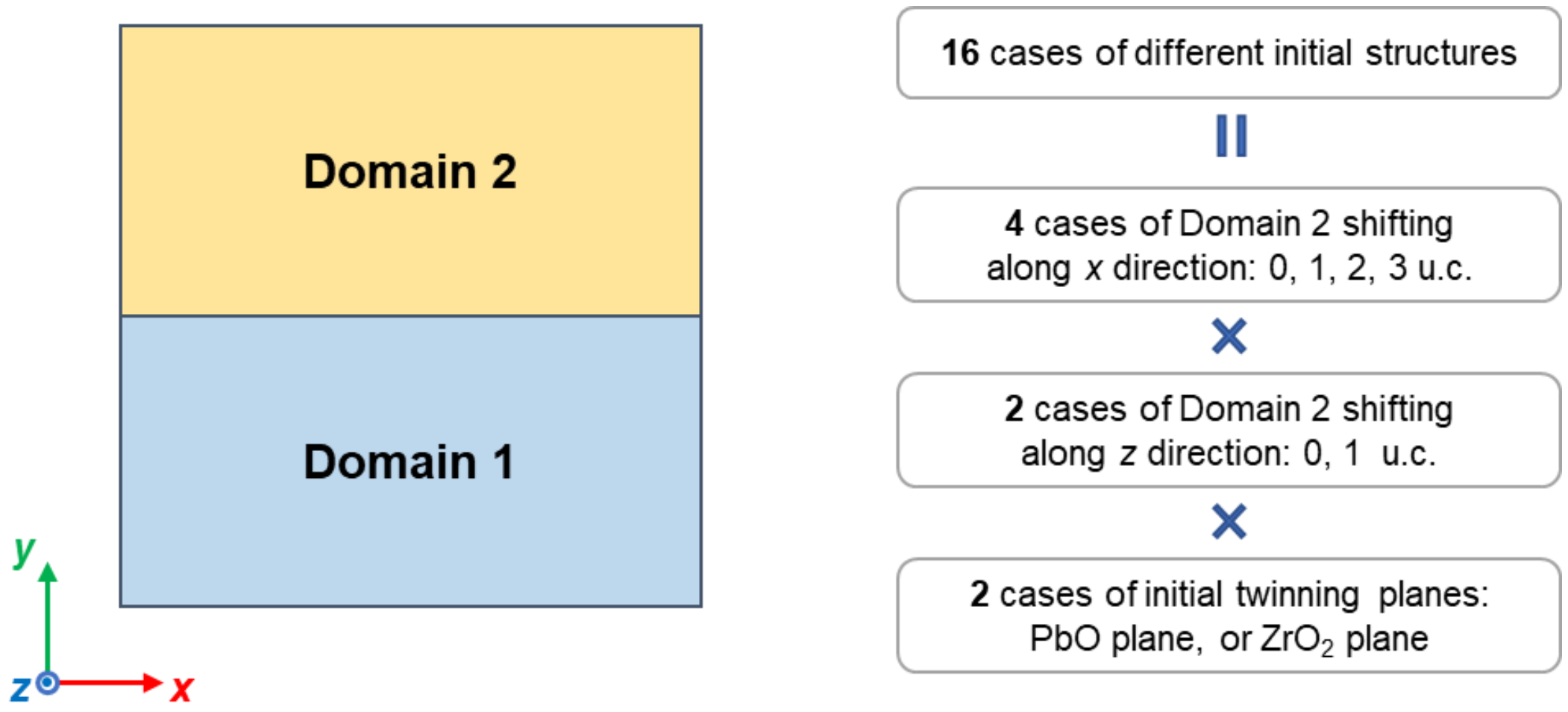

Figure S1. Illustration of the 16 initial models built for the second-principles simulations.

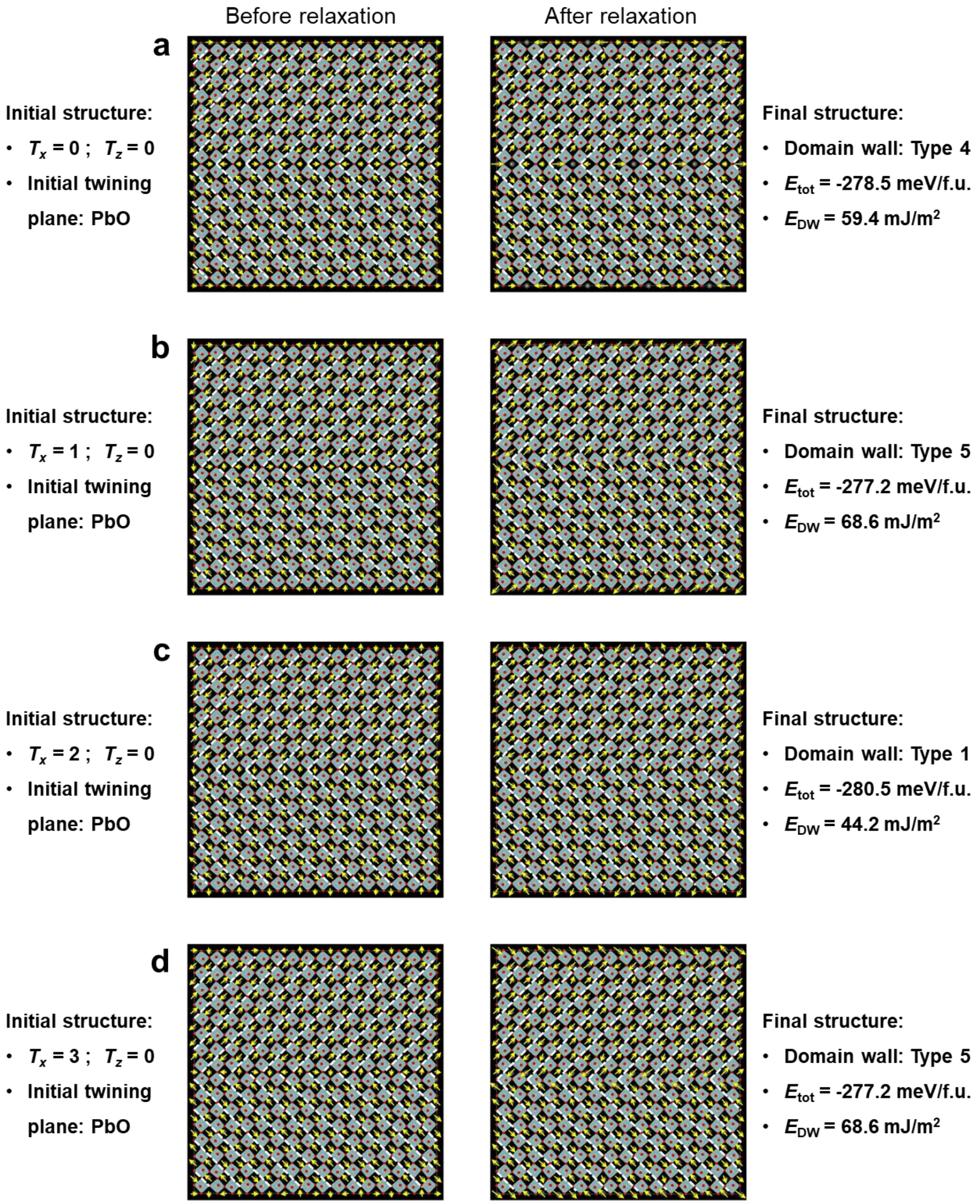

Figure S2. Initial and the corresponding relaxed structures for PbO plane twin boundaries with no shift of Domain 2 along *z* direction ($T_z = 0$). The shift of Domain 2 along *x* is (a) $T_x = 0$ u.c., (b) $T_x = 1$ u.c., (c) $T_x = 2$ u.c., and (d) $T_x = 3$ u.c., respectively.

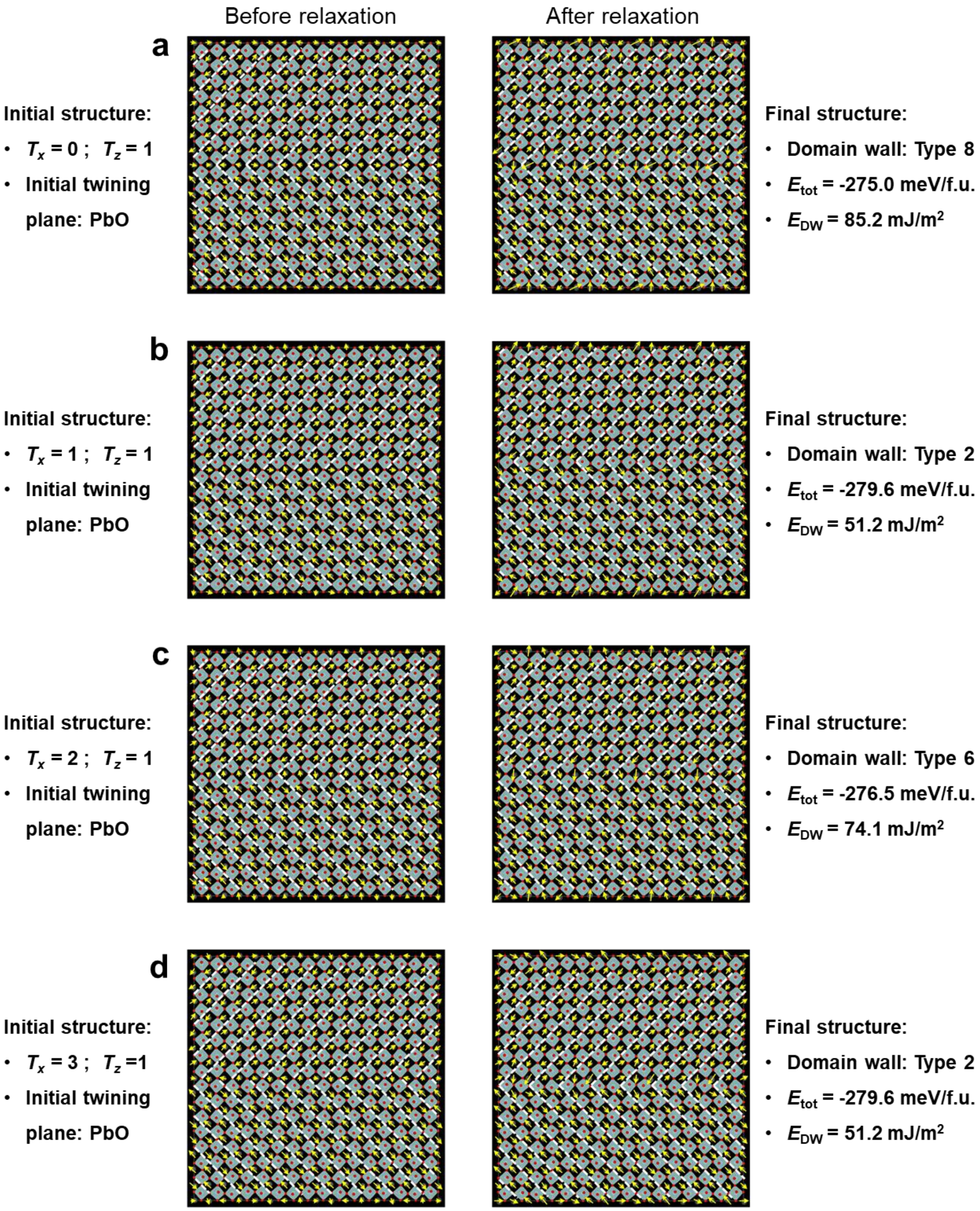

Figure S3. Initial and the corresponding relaxed structures for PbO plane twin boundaries with no shift of Domain 2 along *z* direction ($T_z$ = 1). The shift of Domain 2 along *x* is (a) $T_x$ = 0 u.c., (b) $T_x$ = 1 u.c., (c) $T_x$ = 2 u.c., and (d) $T_x$ = 3 u.c., respectively.

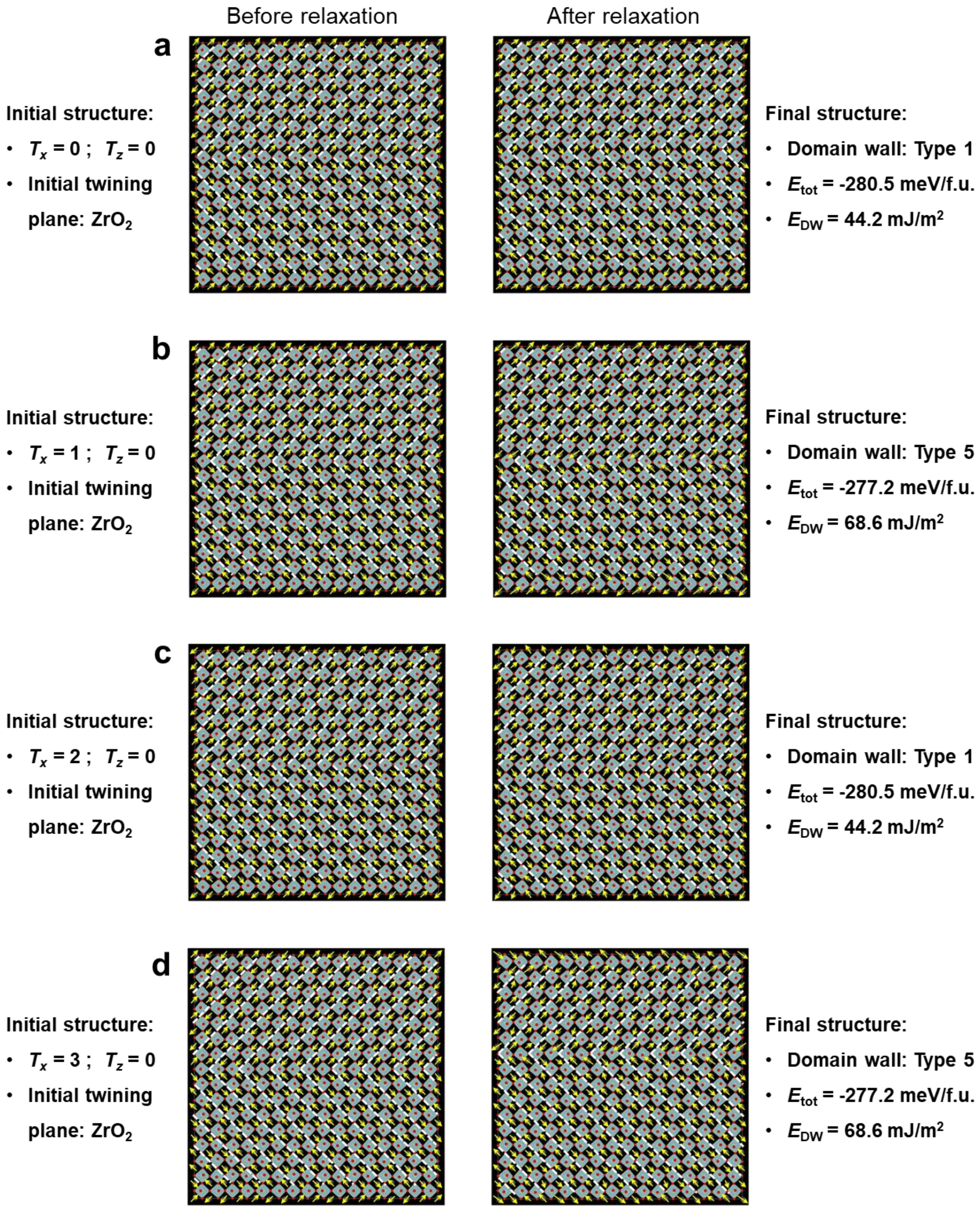

Figure S4. Initial and the corresponding relaxed structures for $ZrO_2$ plane twin boundaries with no shift of Domain 2 along $z$ direction ($T_z = 0$). The shift of Domain 2 along $x$ is (a) $T_x = 0$ u.c., (b) $T_x = 1$ u.c., (c) $T_x = 2$ u.c., and (d) $T_x = 3$ u.c., respectively.

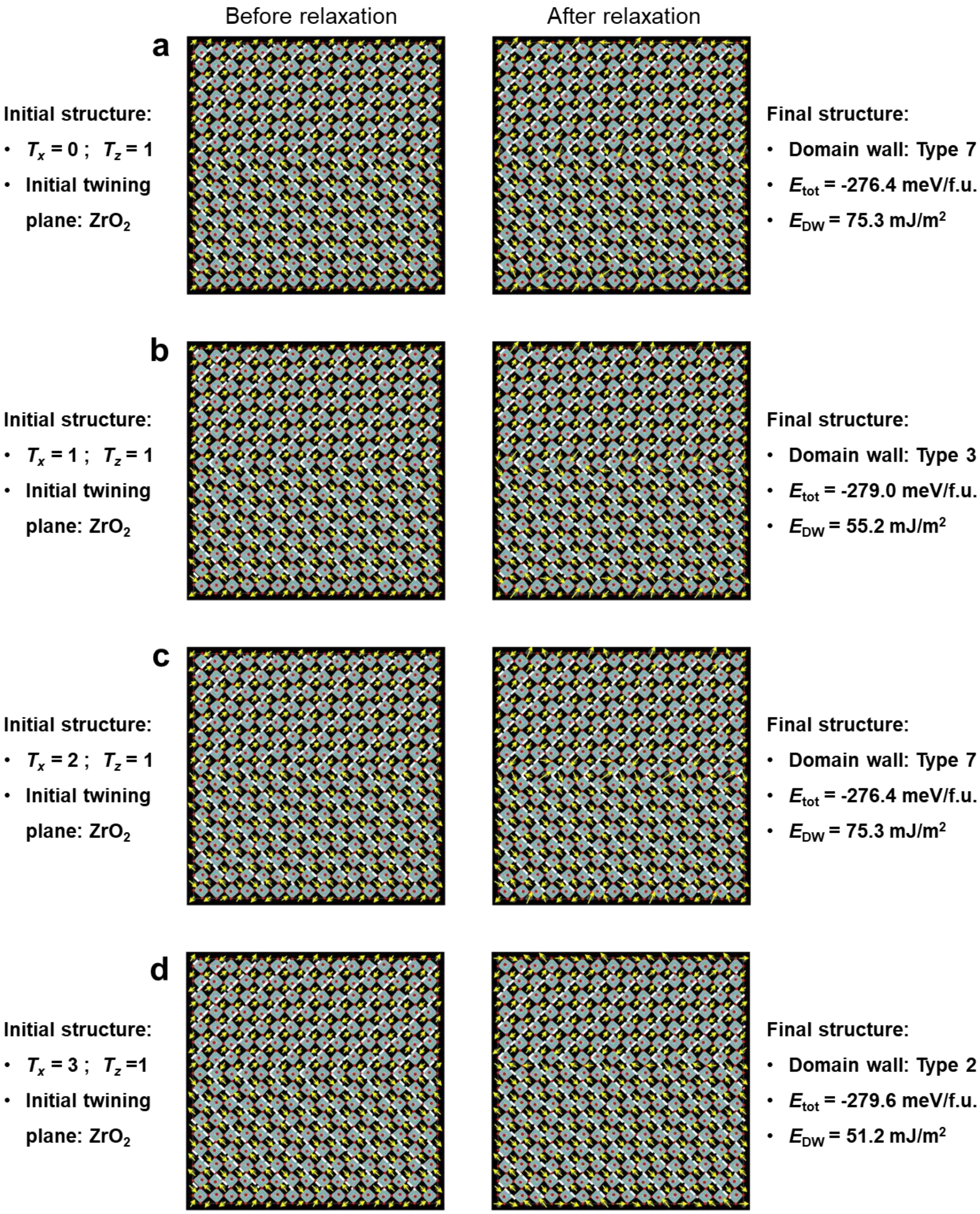

Figure S5. Initial and the corresponding relaxed structures for $ZrO_2$ plane twin boundaries with no shift of Domain 2 along *z* direction ($T_z$ = 1 u.c.). The shift of Domain 2 along *x* is (a) $T_x$ = 0 u.c., (b) $T_x$ = 1 u.c., (c) $T_x$ = 2 u.c., and (d) $T_x$ = 3 u.c., respectively.

Table S2. The second-principles simulation predicted twinning boundaries in PZO and the corresponding domain wall energy $E_{DW}$. The arrows show the local Pb displacements $\delta_{Pb}$, with colours indicating the directions. The intensity of the colours of the squares shows the amplitude of $\varphi_x = \omega_x\left(-1\right)^{i_x+i_y+i_z}$, where $\omega_x$ is the local oxygen octahedral rotation around the $x$-axis (parallel to the wall), $i_x$, $i_y$, $i_z$ are integers locating the oxygen octahedra. The red and blue colours of the squares indicate opposite signs of $\varphi_x$.

| | $E_{DW}$ (mJ/m$^2$) | $\delta_{Pb}$ and $\varphi_x$ near the wall | | $E_{DW}$ (mJ/m$^2$) | $\delta_{Pb}$ and $\varphi_x$ near the wall |
|---|---|---|---|---|---|
| **Type 1** | 44.2 | 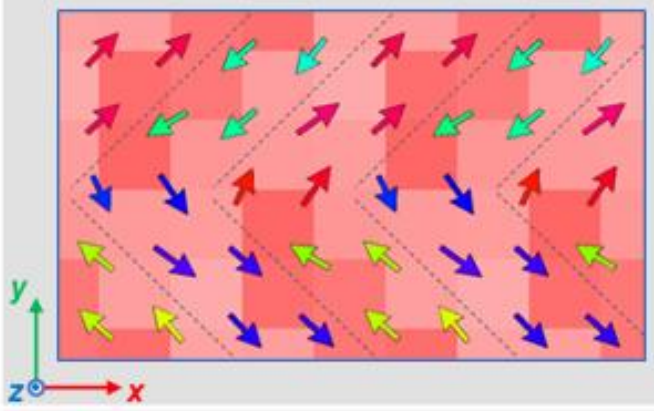 | **Type 5** | 68.6 | 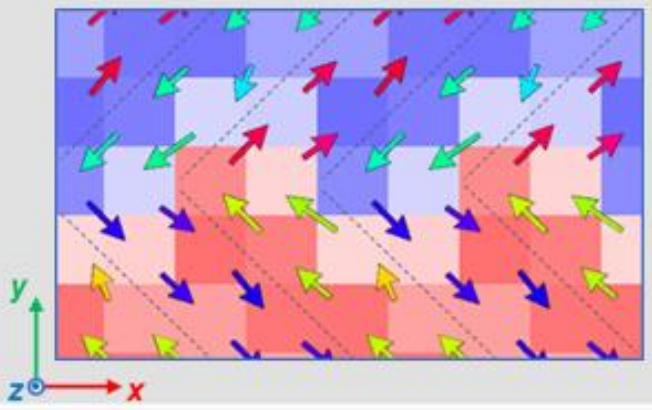 |
| **Type 2** | 51.2 | 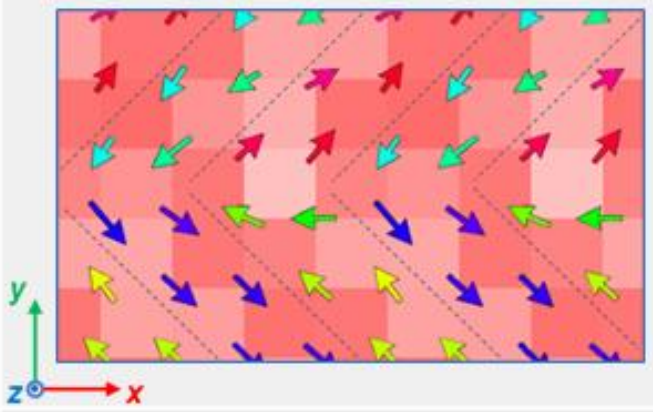 | **Type 6** | 74.1 | 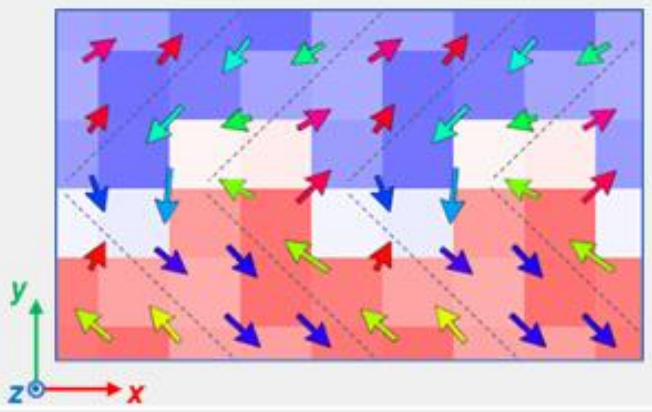 |
| **Type 3** | 55.2 | 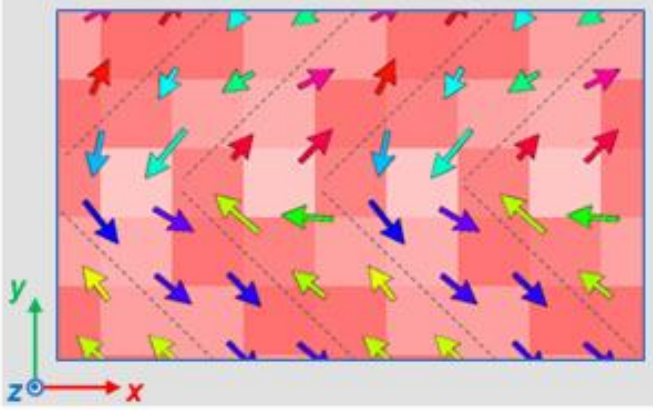 | **Type 7** | 75.3 | 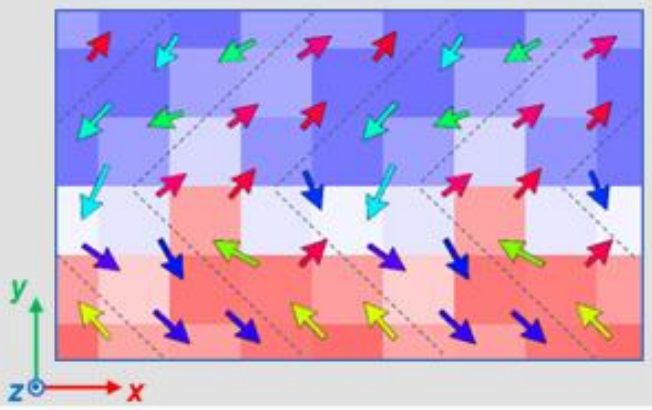 |
| **Type 4** | 59.4 | 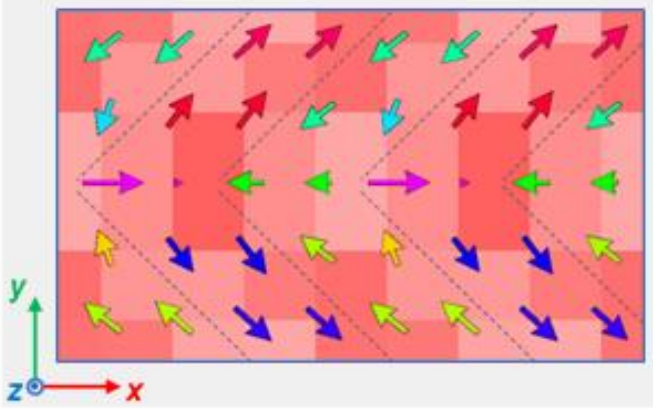 | **Type 8** | 85.2 | 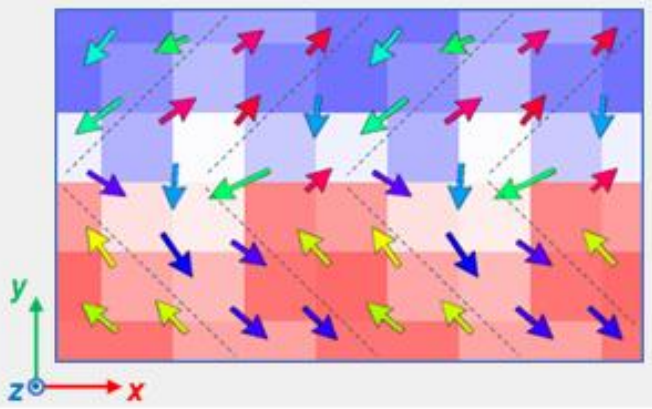 |

**Continuum-field calculation details**

The homogeneous part of the Landau energy (Eq. (1) of the main text) is described in details in Ref. S6. The gradient terms have the following full form:

$$F_{\text{grad}} = \frac{G_{11}}{2}\left[\left(\frac{\partial P_x}{\partial x}\right)^2 + \left(\frac{\partial P_y}{\partial y}\right)^2\right] + G_{12}\frac{\partial P_x}{\partial y}\frac{\partial P_y}{\partial x} + \frac{G_{44}}{2}\left[\left(\frac{\partial P_x}{\partial y}\right)^2 + \left(\frac{\partial P_y}{\partial x}\right)^2\right]$$

$$+\frac{D_{11}}{2}\left[\left(\frac{\partial \phi_x}{\partial x}\right)^2 + \left(\frac{\partial \phi_y}{\partial y}\right)^2\right] + D_{12}\frac{\partial \phi_x}{\partial y}\frac{\partial \phi_y}{\partial x} + \frac{D_{44}}{2}\left[\left(\frac{\partial \phi_x}{\partial y}\right)^2 + \left(\frac{\partial \phi_y}{\partial x}\right)^2\right]$$

$$-W_{1111}\left(P_x\frac{\partial \phi_x}{\partial x}\phi_x + P_y\frac{\partial \phi_y}{\partial y}\phi_y\right) - W_{1212}\left(P_x\frac{\partial \phi_x}{\partial y}\phi_y + P_y\frac{\partial \phi_y}{\partial x}\phi_x\right)$$

$$-W_{1122}\left(P_x\frac{\partial \phi_y}{\partial x}\phi_y + P_y\frac{\partial \phi_x}{\partial y}\phi_x\right) - W_{1221}\left(P_x\frac{\partial \phi_y}{\partial y}\phi_x + P_y\frac{\partial \phi_x}{\partial x}\phi_y\right)$$

The used numerical values of the tensor components are:

| | | | | | |
|---|---|---|---|---|---|
| $G_{11}$ | $5 \times 10^{-3}$ | $D_{11}$ | $0.09 \times 10^{-3}$ | $W_{1111}$ | $6.26 \times 10^{-4}$ |
| $G_{12}$ | $6.54 \times 10^{-3}$ | $D_{12}$ | $-0.71 \times 10^{-3}$ | $W_{1212}$ | $-2.72 \times 10^{-4}$ |
| $G_{44}$ | $2.69 \times 10^{-3}$ | $D_{44}$ | $1.74 \times 10^{-3}$ | $W_{1122}$ | $-1.18 \times 10^{-4}$ |
| | | | | $W_{1221}$ | $-1.97 \times 10^{-4}$ |

For the calculations of the distribution of the fields $\mathbf{P}(x,y)$, $\boldsymbol{\phi}(x,y)$ at the twin boundary, we minimize the energy by using the finite-element methods solver implemented in Wolfram Mathematica. The calculations were done in the rectangular region of size $8a_0 \times 16a_0$ with imposed periodic boundary conditions, the mesh size is $\frac{a_0}{8}$. We start our calculations from the initial distribution of fields reported in [1] inside two domains of size $8a_0 \times 8a_0$, with one domain rotated by 90° and translated in $x$-direction with respect to the other to form the required twin boundary. We consider four initial configurations corresponding to the four classes of the twin boundaries: "easy" head-to-tail, "easy" head-to-head, "hard" head-to-tail, and "hard" head-to-head, where "easy" and "hard" refers to the relative orientations of the average oxygen octahedral tilts of the two domains, and head-to-tail and head-to-head refers to the relative orientation of local polarization at the boundary. The fields are then varied with the local change rate proportional to the variational derivatives of the energy, $\dot{\mathbf{P}} \propto -\frac{\delta F}{\delta \mathbf{P}}$, $\dot{\boldsymbol{\phi}} \propto -\frac{\delta F}{\delta \boldsymbol{\phi}}$, until the distributions of the fields $\mathbf{P}(x,y)$, $\boldsymbol{\phi}(x,y)$ are converged to the stable solution corresponding to $\frac{\delta F}{\delta \mathbf{P}} = 0$ and $\frac{\delta F}{\delta \boldsymbol{\phi}} = 0$. Head-to-tail boundaries converge to their stable solutions. Head-to-head boundaries are metastable, ultimately converging to the head-to-tail configuration; here we report their intermediate state where the rate of change has been diminished but transition towards the energetically more favourable head-to-tail configuration has not happened yet. The results of the calculations, in addition to the ones reported in the main text, are shown in Figs. S1 and S2.

The observation of vorticity and antivorticity can be reproduced in a simplified continuum-field model only consisting of two ingredients: the homogeneous part of the Landau energy depending on $\mathbf{P} = (P_x, P_y)$, and the polarization correlation gradient term. These ingredients are not enough to stabilise the modulated phase corresponding to ↑↑↓↓ polar structure of PZO; but feeding the same initial distributions of the polar fields that we used for to head-to-tail and head-to-head twin boundaries calculations, we report well defined vortices at the head-to-tail and antivortices at the head-to-head boundary during the early stages of the polarization damped relaxation – see Fig. S3. Analogous results are obtained when only considering the isotropic part of the polarisation correlation tensor ($G_{12} = 0, G_{11} = G_{44} = 2.69 \times 10^{-3}$).

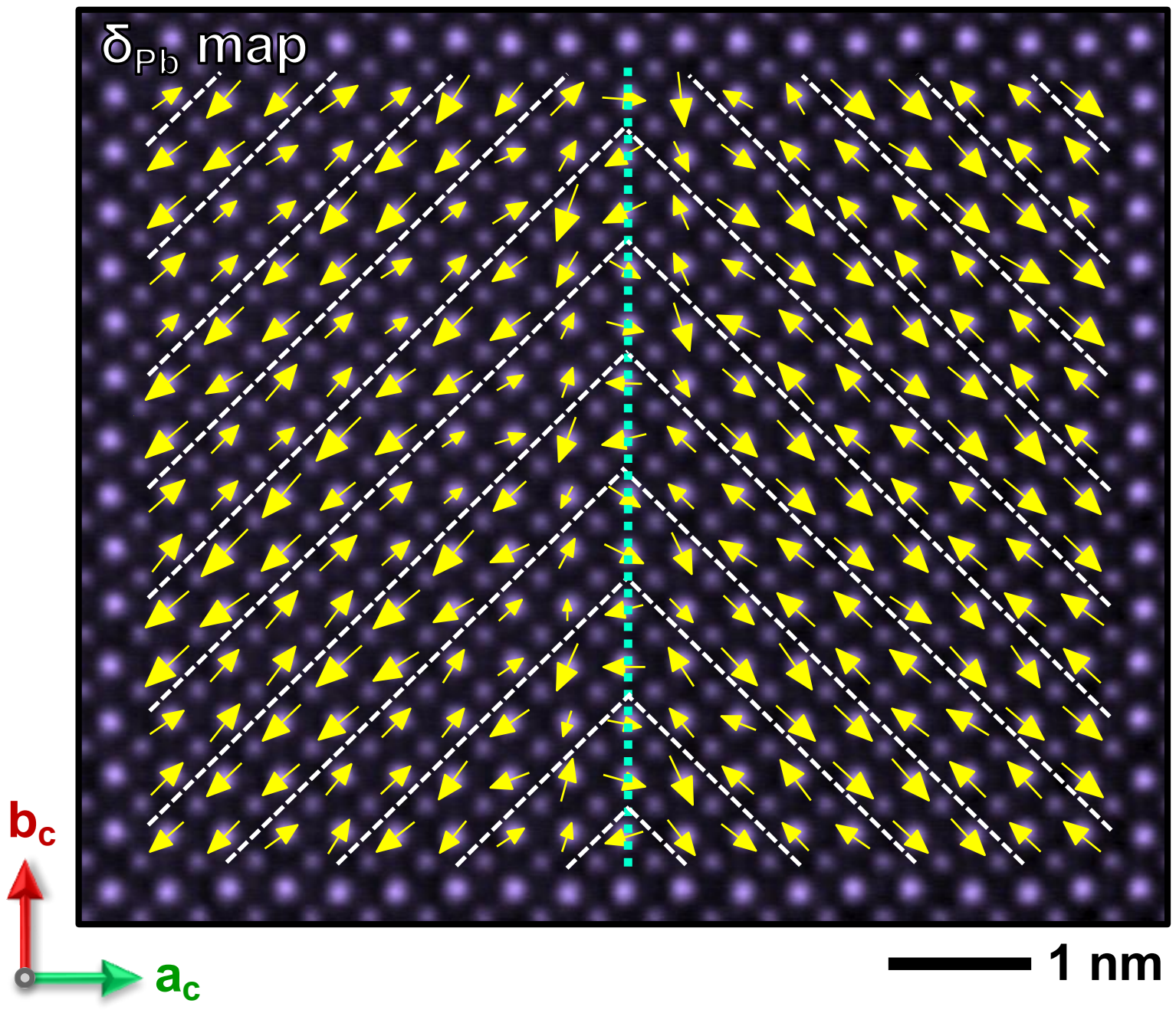

Figure S6. STEM-HAADF image acquired from straight twinning boundaries with the twinning boundary at PbO plane. The HAADF image is overlaid by corresponding Pb displacement maps. Head-to-tail Pb displacement configurations are evident.

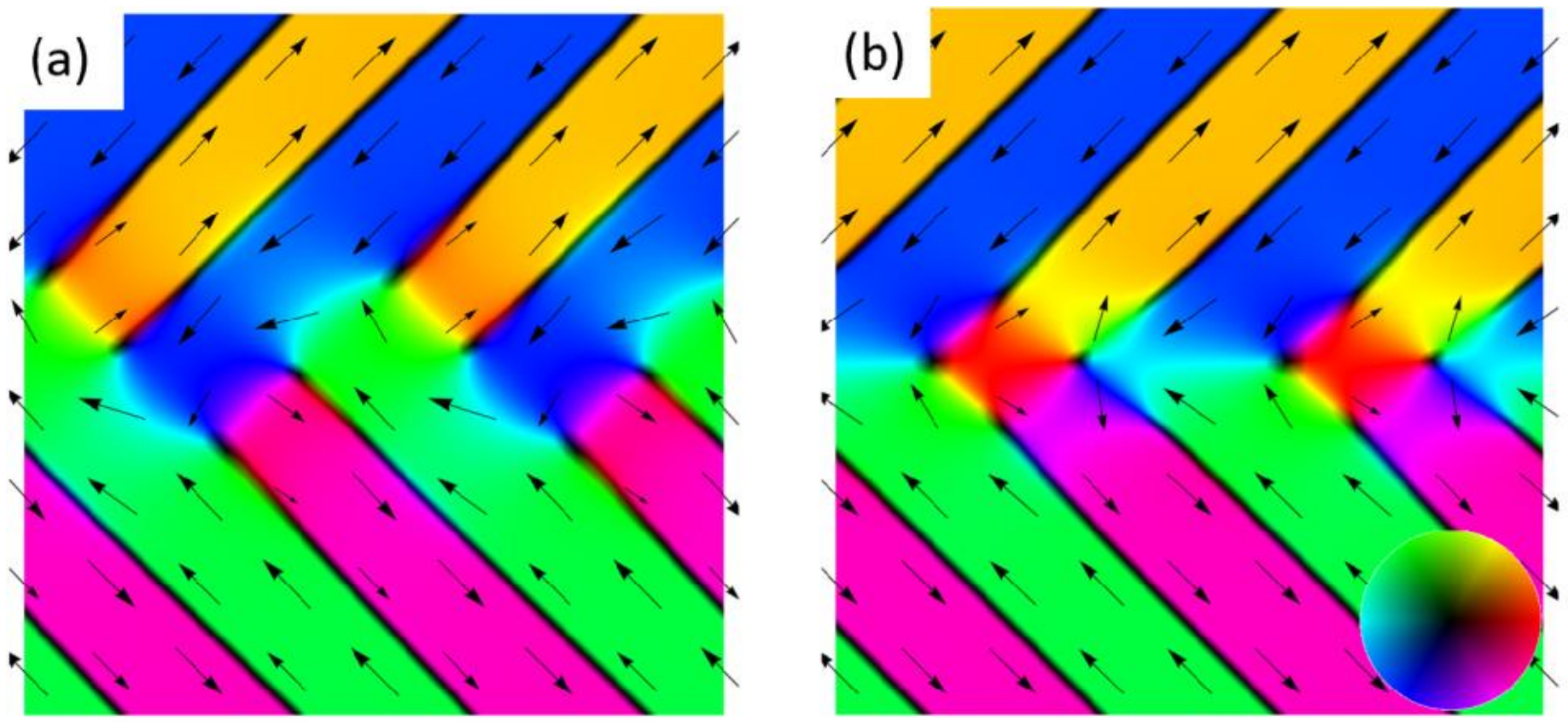

Figure S7. Distribution of polarisation $\mathbf{P}(x, y)$ obtained in the continuum field study of "hard"-type twin boundaries: head-to-tail (a) and head-to-head (b). Colour hue and intensity correspond to, respectively, local orientation and magnitude of polarisation. The highest colour intensity corresponds to $\sim$30% of maximum polarisation obtained in the calculations. Inset: colour hue-intensity wheel.

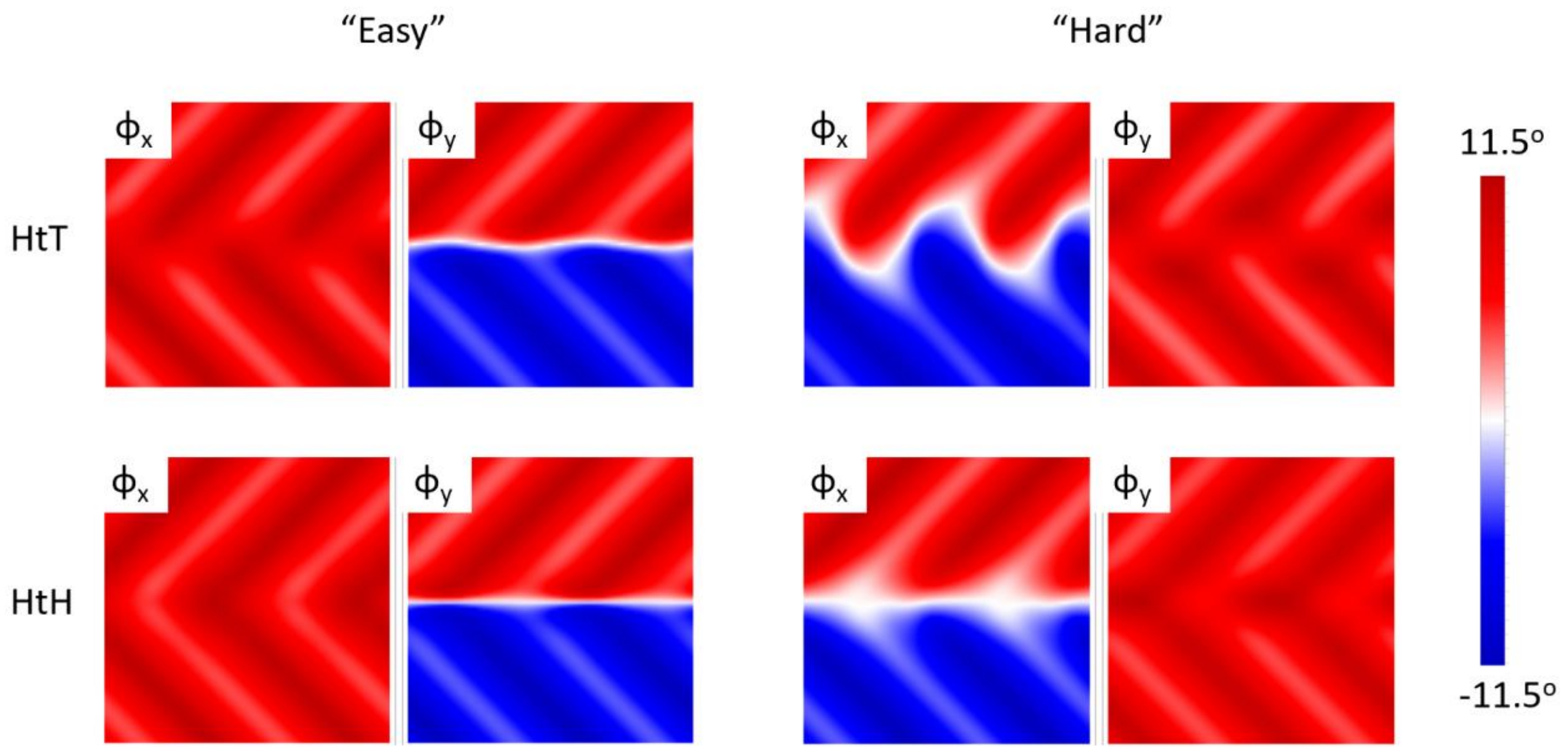

Figure S8. Distribution of tilt $\boldsymbol{\phi}(x, y)$ obtained in the continuum field study of "easy"- and "hard"-type twin boundaries: head-to-tail (HtT) and head-to-head (HtH).

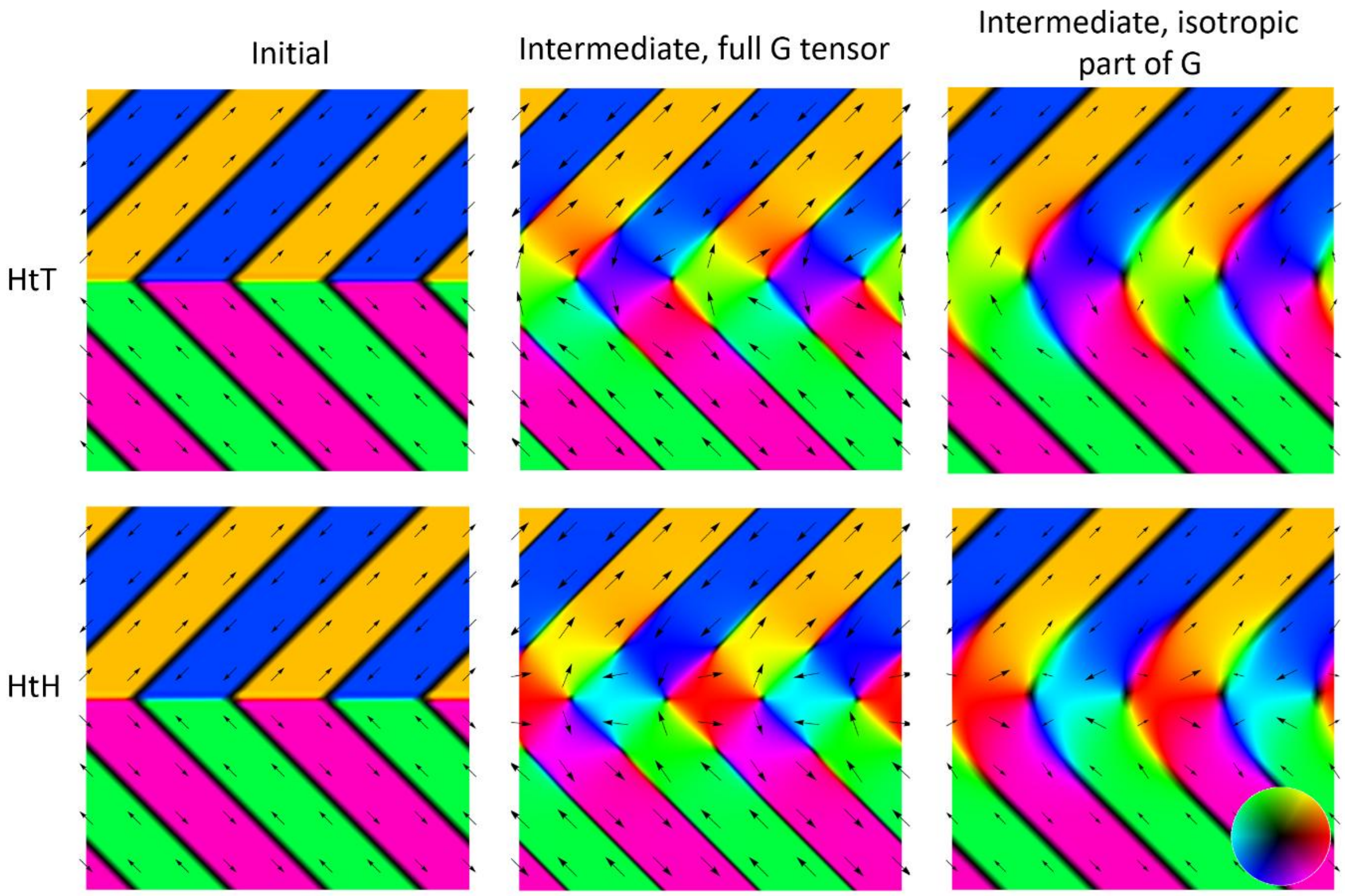

Figure S9. Intermediate distribution of polarisation $\mathbf{P}(x, y)$ obtained in the continuum-field damped relaxation study based only on polarisation-dependent energy terms. The highest colour intensity corresponds to ~30% of maximum polarisation obtained in the calculations. Inset: colour hue-intensity wheel.

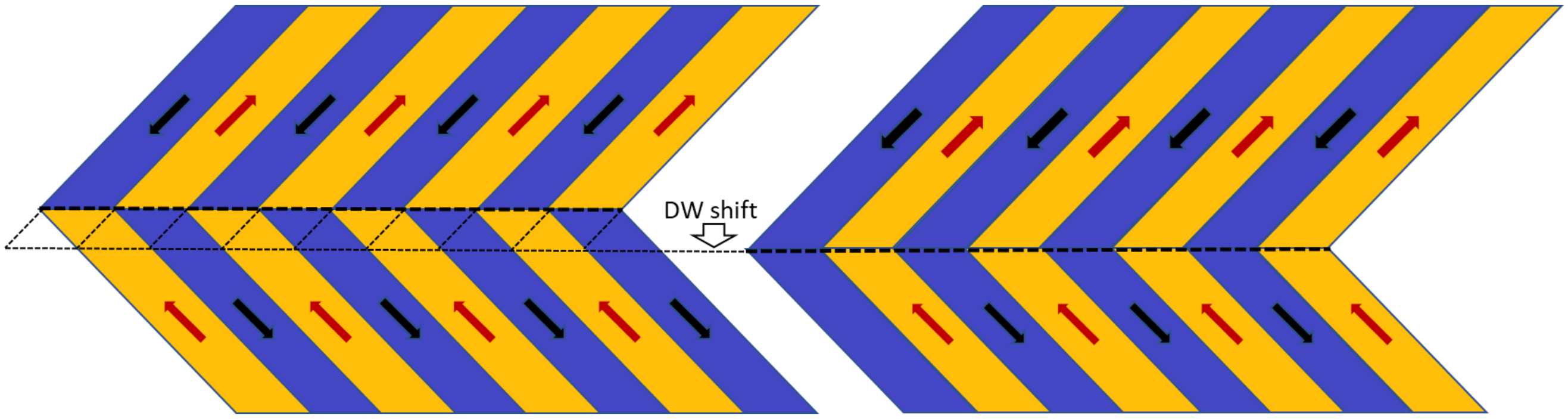

Figure S10. Schematic of ferroelastic domain wall motion (DW shift) in an antiferroelectric. As the DW moves down by one unit cell, geometry dictates that the “herringbone” antipolar order must change from head-to-head to head-to-tail. Obviously, geometry works both ways, and, as the DW keeps moving, the configuration will alternate back and forth between head-to-head and tail-to-tail.

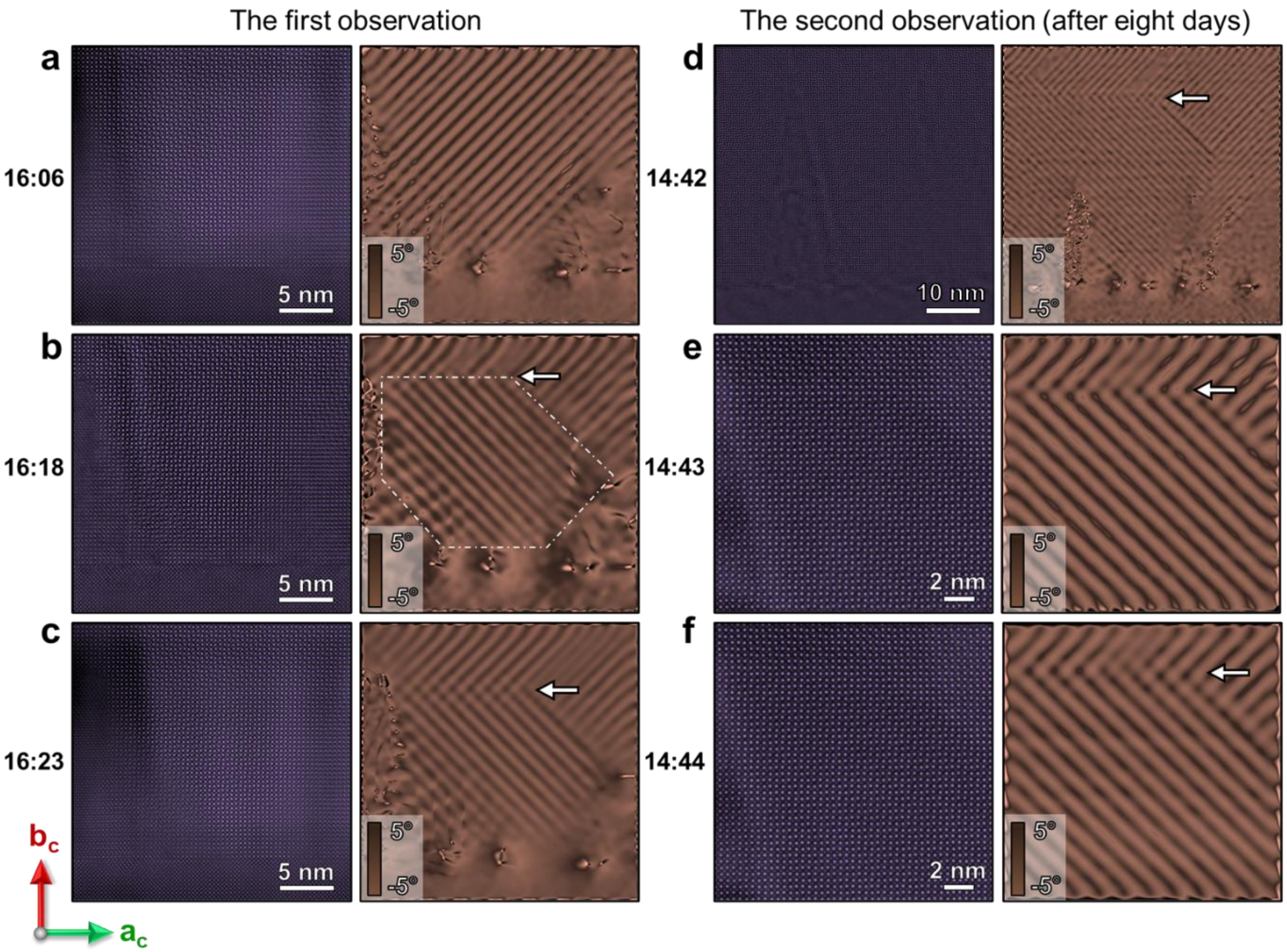

Figure S11. Observation of twinning boundary evolution at different times. Figures 3a, 3b and 3c in the main text are extracted from the twinning boundary part in panels (b), (e) and (f), respectively, as indicated by white arrows.

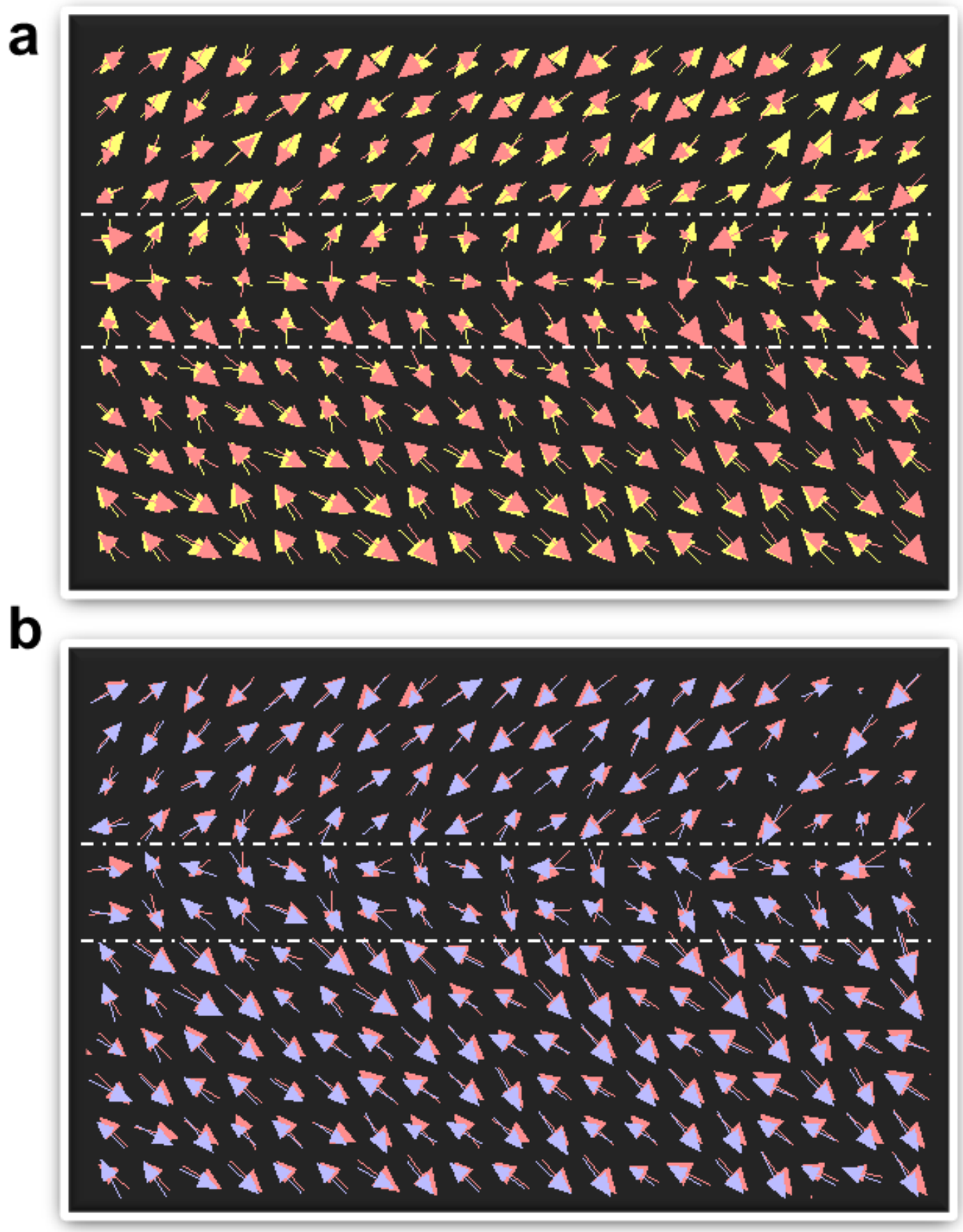

Figure S12. (a) The overlaid Pb displacement map combines Figures 3a (with yellow arrows) and 3b (with pink arrows). (b) The overlaid Pb displacement map merges Figures 3b (with pink arrows) and 3c (with light purple arrows).